\documentclass[12pt]{article}

\usepackage{amsmath,amsfonts,amssymb,bm}

\usepackage[utf8]{inputenc}
\usepackage{amsmath,amssymb}
\usepackage{latexsym}
\usepackage{slashed}
\usepackage{authblk}
\usepackage{xcolor}
\usepackage{hyperref}

\usepackage{cancel}
\usepackage{indentfirst}
\usepackage{graphicx}
\usepackage{epstopdf}
\usepackage{dcolumn}
\usepackage{bm}
\usepackage{lipsum}
\usepackage{slashed}
\usepackage{enumitem}
\usepackage{txfonts}
\usepackage{esint}
\usepackage{xcolor}
\usepackage{mathtools}
\usepackage{transparent}
\usepackage{upgreek}
\usepackage{tabularx}
\usepackage{csquotes}

\begin{document}
\title{Space-like pion off-shell form factors in the Bethe-Salpeter approach}

\author{ 
	\parbox{\linewidth}{\centering
          S. G.~Bondarenko$^{1,2}$,
          M. K.~Slautin$^{1,2}$\thanks{E-mail:~slautin@jinr.ru}
}
}

\date{}

\maketitle


\begin{center}
{
$^1$ \it Bogoliubov Laboratory of Theoretical Physics, JINR, Dubna, 141980 Russia\\
$^2$ \it Dubna State University, Dubna, 141980 Russia
}
\end{center}
 

\maketitle
\begin{abstract}
In the paper, the off-shell electromagnetic pion form factors in the Bethe-Salpeter formalism are
considered. The separable kernel of the first rank quark-antiquark interaction is used to solve the
equation analytically. The semi-off-shell pion form factors \( F_1 \) and \( F_2 \), which are related to each other
by the Ward-Takahashi identity, are calculated. The obtained off-shell form factors as well as static
properties of the pion are compared with the results of other authors.
\end{abstract}


\section{Introduction}

The pion, the simplest coupled system of a quark and antiquark, occupies a special place among the mesons. Its relatively low mass (much smaller than that of other mesons) makes it a key element in describing nuclear dynamical processes. Currently there are a number of different models for the theoretical description of the pion:
QCD sum rules~\cite{Nesterenko:1982gc};  nonrelativistic potential model~\cite{Godfrey:1985xj};
a relativistic model using the light-front formalism~\cite{Jacob:1988as}; the Nambu-Jona-Lasinio models
~\cite{Nambu:1961tp,Zhang:2024dhs,Anikin:1995cf,Bernard:1986ti,Hatsuda:1985ey};
a model based on chiral symmetry~\cite{Gross:1991te}; an instanton pion
model~\cite{Anikin:2000rq}; lattice calculations ~\cite{ExtendedTwistedMass:2023hin},~\cite{Gerardin:2019vio};
recent models based on the Bethe-Salpeter equation with
dressed quark and gluon propagators~\cite{Maris:2000sk,Kekez:2020vfh,Hernandez-Pinto:2023yin}; the Polyakov-Nambu-Yona-Lasinio model~\cite{Friesen:2025}.

The electromagnetic (EM) form factor of the pion
$F_{\pi}(Q^2)$ is determined on the mass shell and characterizes the spatial distribution of charge inside the pion as a function of the square of the transferred 4-momentum ($q^2$). In regions where $q^2$ is small, this form factor can be measured directly in experiments on elastic scattering of pions on electrons. However, direct determination of the form factor is difficult for intermediate and high values of $q^2$ due to the short lifetime of the pion, which makes it unsuitable for such studies.

An alternative approach for studying the form factor in these regions experimentally is based on the exclusive Sullivan process~\cite{Sullivan:1971kd}. In particular, it involves the analysis of the cross sections of the $^1H(e,e' \pi^ +) n$ EM reaction. In this case, however, the virtual pion in the intermediate state means that the measured form factor appears to be half-off-shell. This requires additional theoretical analysis for interpretation.

In this paper, the half-off-shell EM form factors and
off-shell effects of the
pion are investigated using a model based on the relativistic covariant Bethe-Salpeter equation with a separable kernel. 
The choice of this model is due to the simplicity of the analytical solution 
of the vertex function of the pion.
The Ward-Takahashi identity~\cite{Ward:1950xp},~\cite{Takahashi:1957xn} is verified for the off-shell pion form factors $F_1$ and $F_2$. The obtained physical constants and pion form factors are also compared with the results of previous studies~\cite{Choi:2019nvk},~\cite{Leao:2024agy}.

The paper is organized as follows:
in Section 2 the basic formulae of the formalism are given, 
in Section 3 the off-shell form factors $F_1(Q^2,t)$ and $F_2(Q^2,t)$ are determined, and the Ward-Takahashi identity is discussed.
In Section 4 a method of computing form factors is described, 
in Section 5 the obtained results are discussed, 
and in Section 6 the main results and conclusions are summarized.

\section{Bethe-Salpeter approach with a separable kernel}
The Bethe-Salpeter equation for the vertex function of the pion is written as follows~\cite{Ito:1991sz}:
\begin{equation}
\Gamma_{\alpha\beta}(k;p)=i\int\dfrac{d^4k''}{(2\pi)^4}V_{\alpha\beta:\epsilon\lambda}(k,k'';p)S_{\lambda\eta}(k''+p/2)\Gamma_{\eta\zeta}(k'';p)S_{\zeta\epsilon}(k''-p/2).
\label{eqgamma}
\end{equation}
where $p,k$ are the total and relative 4-momenta of the pion, respectively ($p=k_1+k_2$ and $k=(k_1-k_2)/2$), and $V(k',k;p)$ is the interaction kernel. 
The pion mass $m_{\pi}$ is defined on the pion mass shell as $p^2=m_\pi^2$ where $m_\pi=140$ MeV.
The quark propagator with mass $m$ has the form
$S(k)=(\hat{k}-m+i\epsilon)^{-1}$. The Greek symbols denote the Dirac indices.

In this paper, a separable interaction kernel of rank one is considered
in the following form:
 \begin{equation}
     V_{\alpha\beta:\delta\gamma}(k',k;p)=\gamma_{\alpha\beta}^5 f(k'^2) \times \gamma_{\delta\gamma}^5 f(k^2).
 \end{equation}

For simplicity, in the paper only the $k^2$ dependence of the scalar function is considered. In this case the solution for the vertex function~(\ref{eqgamma}) of the charged pion can be written in the following form:
\begin{equation}
    \Gamma(k;p)\equiv\Gamma(k)={N\gamma^5}{f(k^2)},
\end{equation}
where $N$ is the normalization constant.

The radial part of the vertex function is chosen in monopole form:
\begin{equation}
f(k^2)=\dfrac{1}{k^2-\Lambda^2+i\epsilon},
\end{equation}
where the parameter $\Lambda$ is related to the size of the pion.


\section{Off-shell electromagnetic form factors of the pion}

The photon-pion vertex $G_\mu$ in the general case can be represented in the following form~\cite{Rudy:1994qb}:
\begin{equation}
    G_\mu(p,p')=(p'+p)_{\mu}F_1(q^2,p^2,p'^{2})+q_{\mu}F_2(q^2,p^2,p'^{2}),
\label{gammapi}
\end{equation}
where $p, p'$ are the initial and final 4-momenta, $q=p'-p$ is the transferred 4-momentum
of the virtual photon at the vertex. The Ward-Takahashi identity is satisfied by this vertex on the off-shell~\cite{Rudy:1994qb},~\cite{Itzykson:1980rh}:
\begin{equation}\label{WTI}
    q^{\mu}G_{\mu}(p,p')=\Delta^{-1}(p')-\Delta^{-1}(p),
\end{equation}
where
\begin{equation}
    \Delta(p)=\dfrac{1}{p^2-m_{\pi}^2-\Pi(p^2)+i\varepsilon}
\end{equation}
is a fully renormalized propagator, and the renormalized
self-energy of the pion $\Pi(p^ 2)$ is bounded by the condition on the mass shell: $\Pi(m_{\pi}^ 2)=0$.

In the case when the initial pion is off-shell,  
and the final pion is on the mass shell,
it follows from equations~(\ref{gammapi}),(\ref{WTI}) that
\begin{equation}\label{WTIF2}
    F_2(Q^2,t)=\dfrac{t-m_\pi^2}{Q^2}[F_1(0,t)-F_1(Q^2,t)],
\end{equation}
where the square of the transferred momentum has the space-like form $Q^2=-q^2$,
$t=p^2$ is the off-shell parameter.
The dependence on $p'^2=m_\pi^2$ is omitted for the half off-shell form factors below.

It should be noted that when the pion is completely on the mass 
shell ($p^2=p'^2=m_\pi^2$), then $F_2(Q^2,m_\pi^2)=0$,
and it gives no contribution to the normalization of the form factors $F_1(0,m_\pi^2)=1$. 
It also ensures EM current conservation.

The function $g$ can be determined using equation~(\ref{WTIF2})
\begin{equation}\label{WTIg}
    g(Q^2,t)\equiv \dfrac{F_2(Q^2,t)}{t-m_\pi^2},
\end{equation}
which does not turn to zero on the mass shell. 
The value of $g(0,m^2_p)$, as shown in equations (\ref{WTIF2}) and (\ref{WTIg}) is connected to the charge radius of the pion:
\begin{equation}
    g(0,m_\pi^2)=-\dfrac{\partial}{\partial Q^2}F_1(0,m_\pi^2)=\dfrac{1}{6}\langle r_\pi^2\rangle.
\end{equation}

The new form factor $g(Q^2,t)$ is a physical quantity that can be observed in experiments.
\section{Computational details}

The EM off-shell pion form factors consist of two contributions. The first is the relativistic impulse approximation (RIA) and the second one is the contribution of the interaction current (int)~\cite{Bondarenko:2025aep}:

\begin{equation}
    F_{(1,2)}(Q^2,t)=F^{\textrm{RIA}}_{(1,2)}(Q^2,t)+F^{\textrm{int}}_{(1,2)}(Q^2,t).
\label{f1f2dir1}
\end{equation}
The form factor expressions are 4-dimensional integrals with first- and second-order poles on the 4-momentum $k_{\mu}$.
These integrals can be calculated using the Feynman parameterization method.
The following representation can be used to obtain expressions for the off-shell form factors $F_1(Q^2,t)$ and $F_2(Q^2,t)$:
\begin{equation}
\int {d^4k}~{k}^{\mu}~\tilde{F}(k,Q^2,t)\rightarrow (p'+p)^{\mu}\int{d^4k}~\tilde{F_{1}}(k,Q^2,t)~+~q^{\mu}\int{d^4k}~\tilde{F_{2}}(k,Q^2,t).
\label{f1f2dir2}
\end{equation}

\section{Calculation results}

The model parameters were obtained in the previous work~\cite{Bondarenko:2025aep}: constitutive quark mass $m=260$ MeV and parameter $\Lambda=550$ MeV. This set of parameters is optimal for describing the pion constants 
(the weak decay constant $f_{\pi}$,
the two-photon decay width $\Gamma_{\pi^0\xrightarrow[]{}\gamma\gamma}$,
the transition radius $r_{\pi\gamma}$ and the charge radius $<r^2_{\pi}>$)
as well as the elastic and transition form factors of the pion.

Table ~\ref{t1} summarizes the results of the calculations of the pion constants ($\sqrt{\langle r_\pi^2\rangle}, f_\pi, g(0, m_\pi^2)$) in comparison with the results from the articles ~\cite{Choi:2019nvk} and ~\cite{Leao:2024agy} as well as with experimental data. Unlike the CON and SYM models, the model presented in this paper describes the constants well, since the parameters were chosen specifically to describe the static properties of the pion.

\begin{table}[htbp]
\caption{Pion constants}
\begin{center}
    \begin{tabular}{ | l | l | l | l | }
 \hline
     Model &$\sqrt{\langle{r^2_\pi}\rangle}$[fm]&$f_\pi$[MeV]&$g(0,m_\pi^2)[\textrm{GeV}^2]$\\ \hline
      $\Gamma_{\pi}^{(\textrm{CON})}~\cite{Choi:2019nvk}$ & 0.713 $\pm$ 0.013  & --- & 2.18 $\pm$ 0.08\\ \hline
     $\Gamma_{\pi}^{(\textrm{SYM})}~\cite{Leao:2024agy}$ & 0.736 & 92.40 & 2.32 \\ \hline
     model~\cite{Bondarenko:2025aep} & 0.679 & 101.38 & 1.97\\ \hline
     Exp. & 0.672(8) & 92.28(7) & 1.93(5) \\ \hline
    \end{tabular}\label{t1}
\end{center}
\end{table}

\begin{figure}[htb]
\centerline{\includegraphics[width=0.5\linewidth]{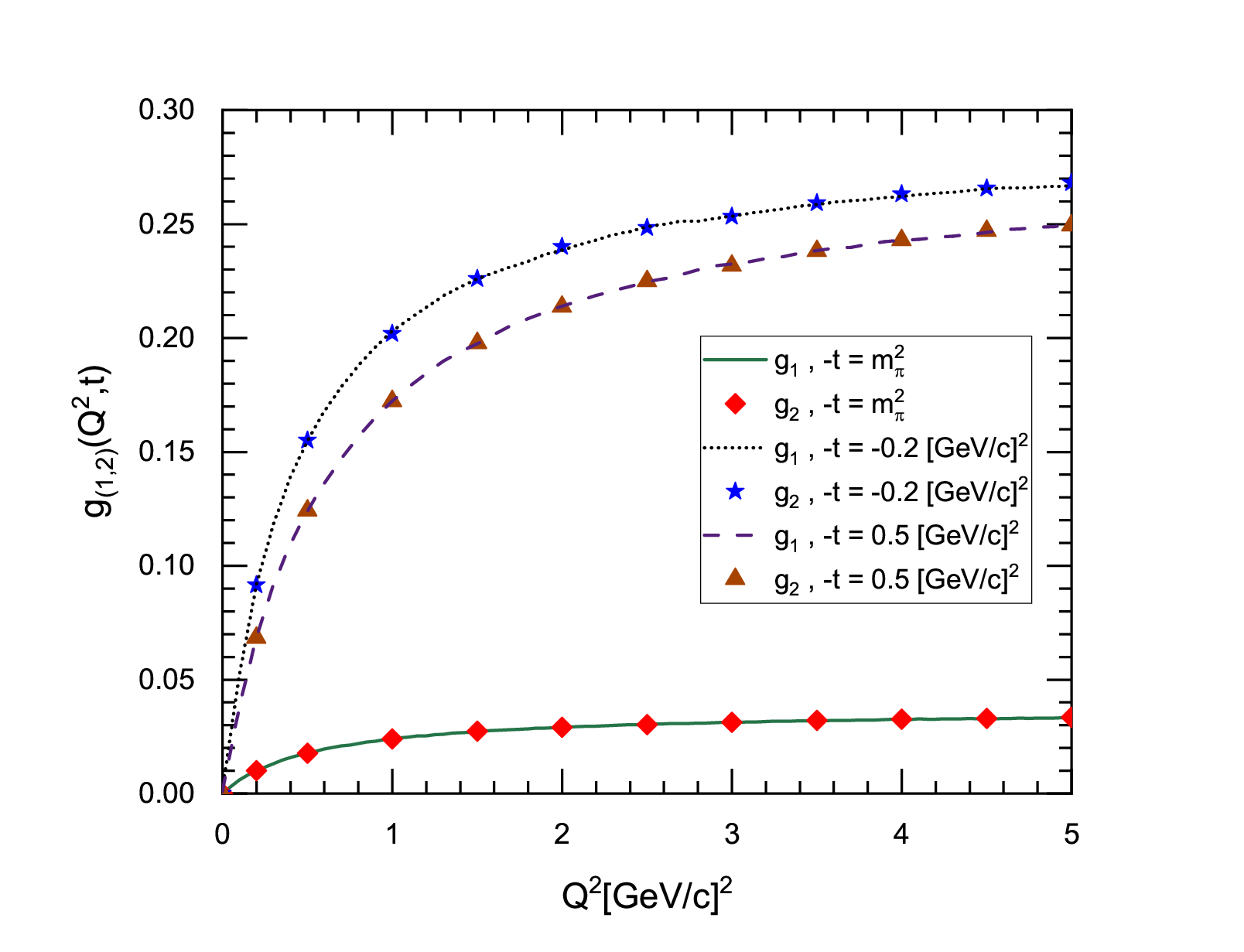}} 
\caption{\label{fig1} The half off-shell form factor $g$ as a function of the square of the transferred momentum $Q^2$ at different values of the off-shell parameter $t$}
\end{figure}

Figure ~\ref{fig1} shows the half-off-shell form factor $g(Q^2,t)$ as a function of the square of the transferred momentum $Q^2$ at different values of the off-shell parameter $t$ ($t=-m_{\pi}^2, t=0.2~\textrm{[GeV/c]}^2, t=-0.5~\textrm{[GeV/c]}^2$). The lines show the
form factors obtained using the Ward-Takahashi identity $g_1=|(t-m_{\pi}^2)[F_1(0,t)-F_1(Q^2,t)]|$, and the symbols show direct calculations using formulas~(\ref{f1f2dir1}),(\ref{f1f2dir2}) ($g_2=|Q^2F_2(Q^2,t)|$). The graph shows that the form factors coincide for any values of $t$ and $Q^2$. It means that the Ward–Takahashi identity is satisfied for the model presented in this paper and that the form factor $F_2(Q^2,t)$ can be expressed in terms of $F_1(Q^2,t)$.

\begin{figure}[htb]
\begin{minipage}[htb]{0.5\linewidth}
{\includegraphics[width=\linewidth]{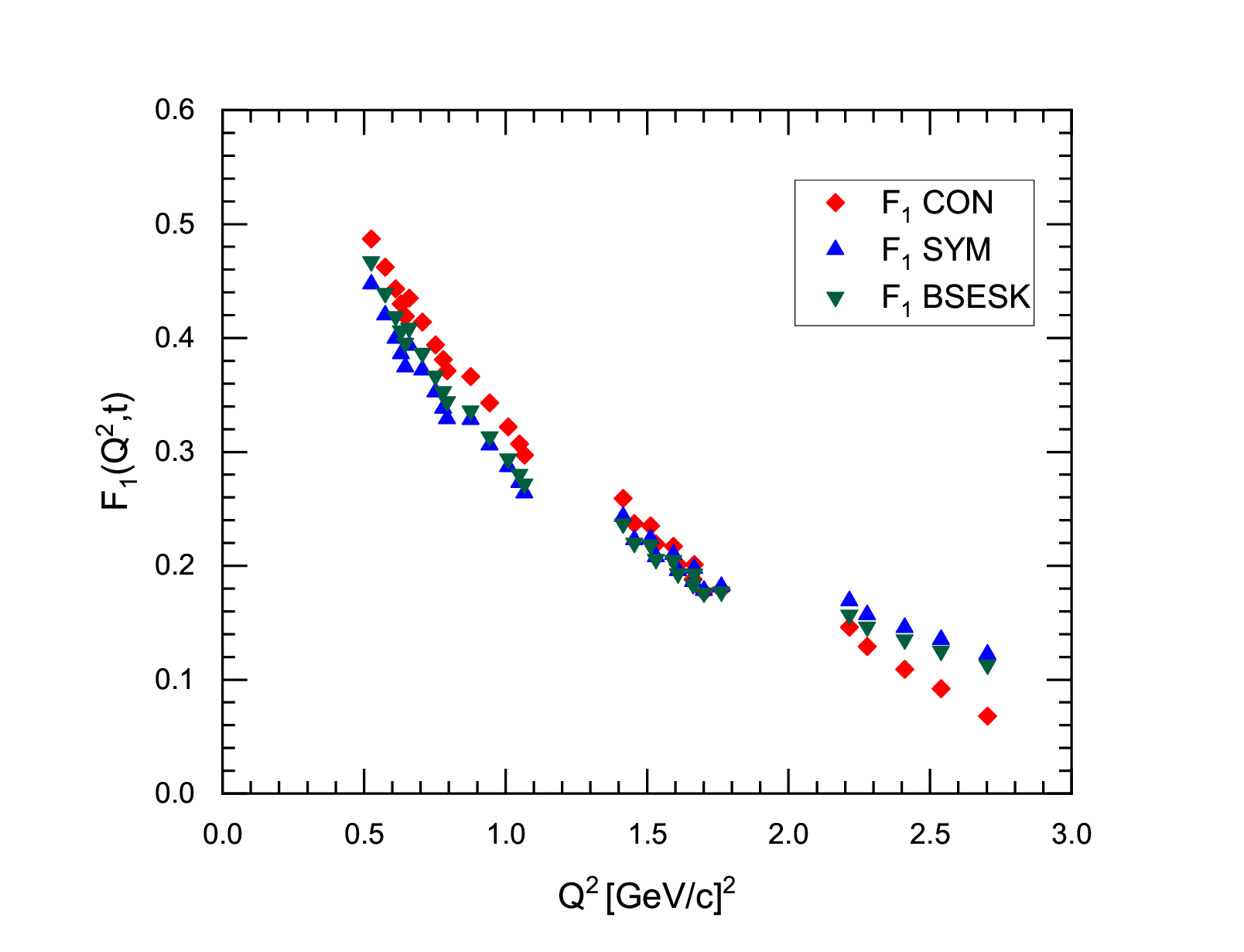}}
\end{minipage}
\begin{minipage}[htb]{0.5\linewidth}
{\includegraphics[width=\linewidth]{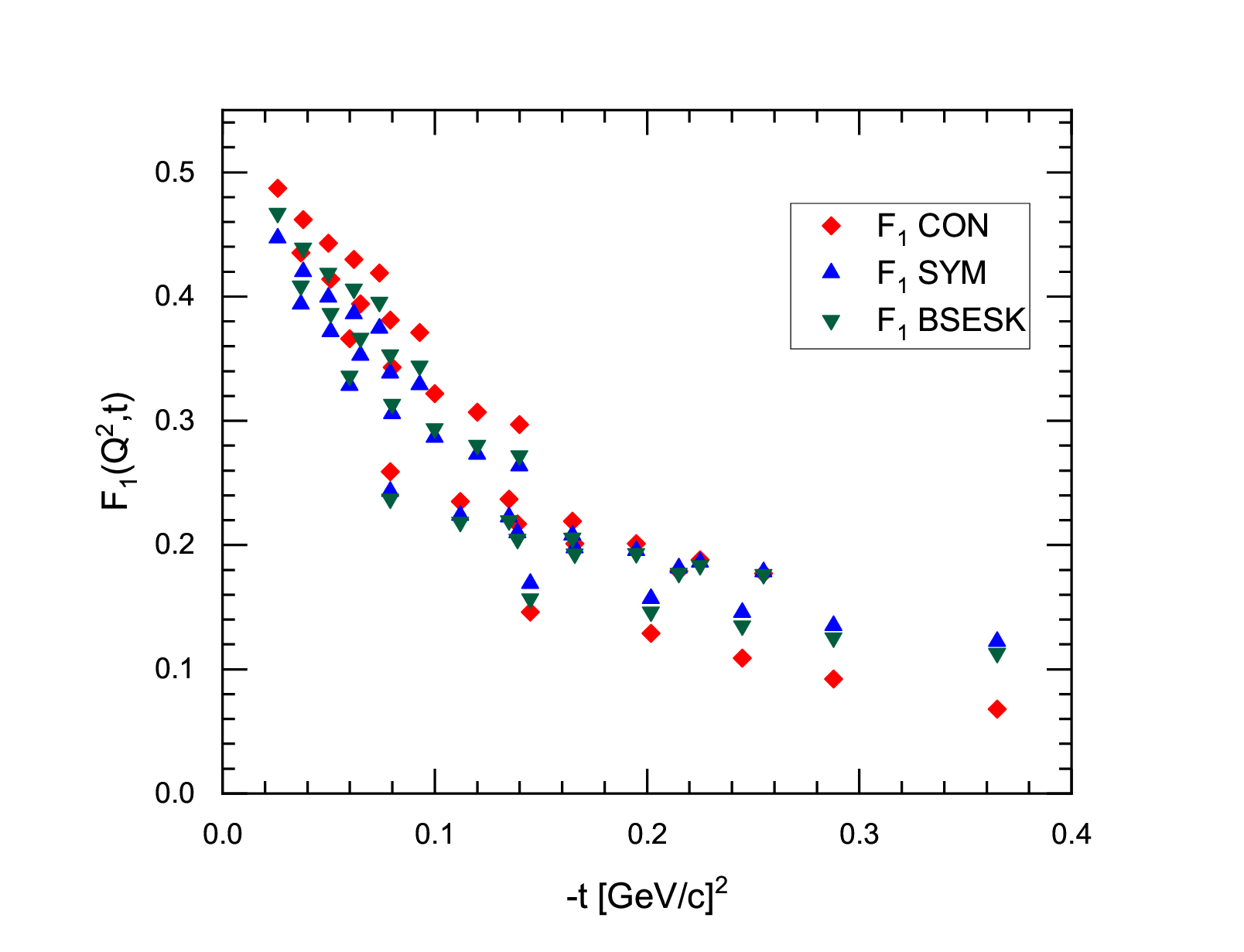}}
\end{minipage}
\caption{\label{fig2}
The half off-shell form factor $F_1(Q^2,t)$ as a function of $Q^2$ (left panel) and $t$ (right panel)}
\end{figure}

\begin{figure}[htb]
\begin{minipage}[htb]{0.5\linewidth}
{\includegraphics[width=\linewidth]{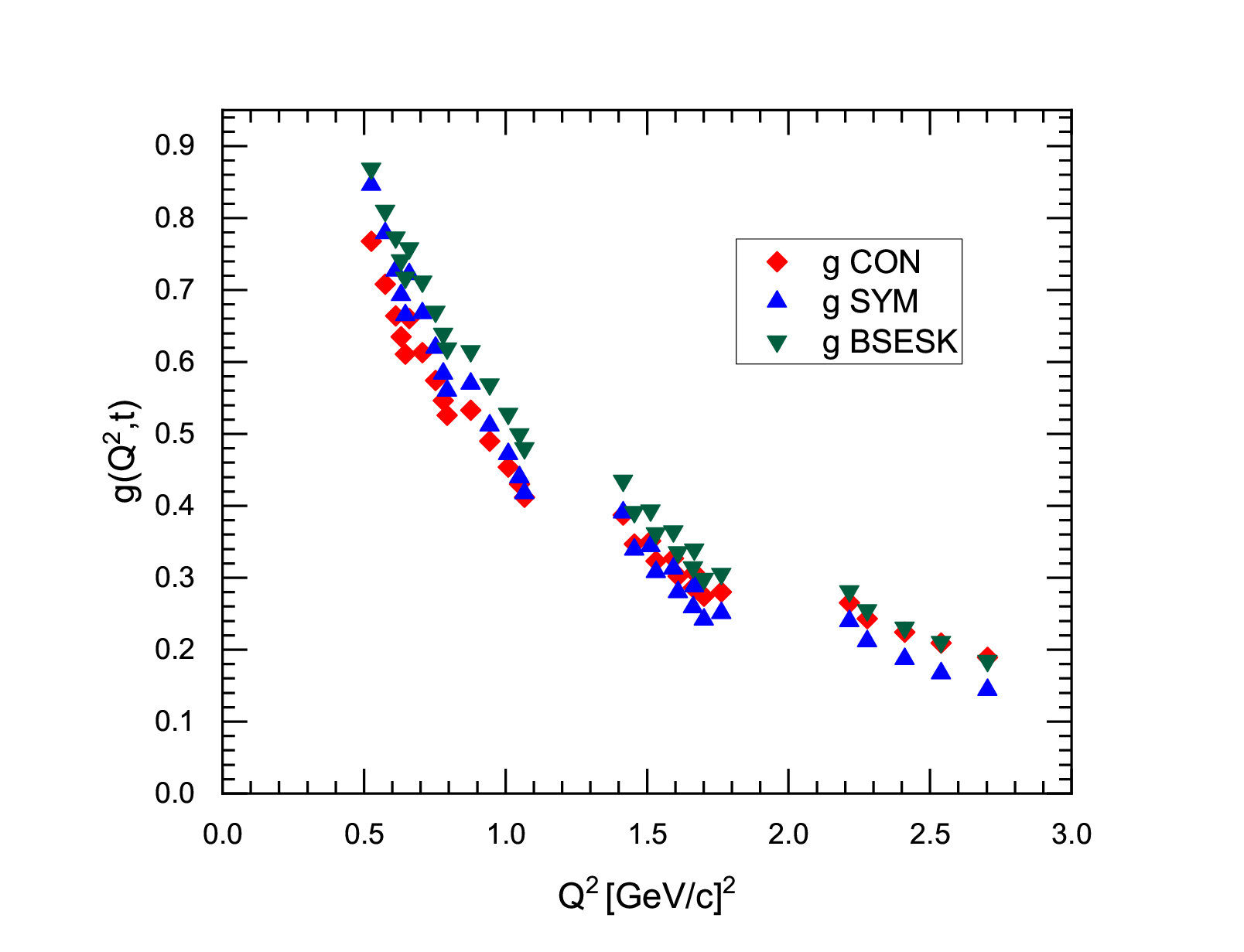}}
\end{minipage}
\begin{minipage}[htb]{0.5\linewidth}
{\includegraphics[width=\linewidth]{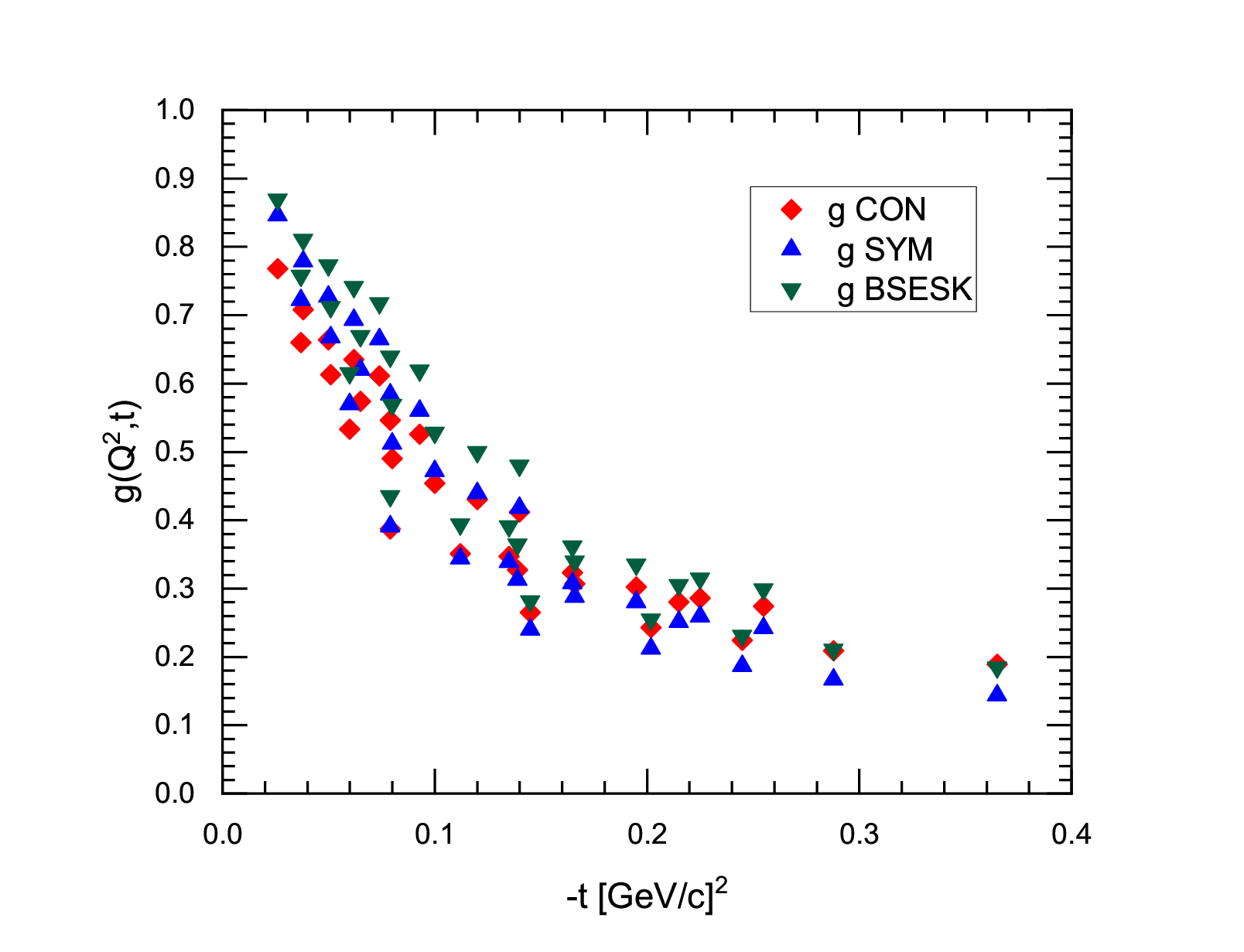}}
\end{minipage}
\caption{\label{fig3}
The half off-shell form factor $g(Q^2,t)$ as a function of $Q^2$ (left panel) and $t$ (right panel)}
\end{figure}

Figure ~\ref{fig2} compares the calculated off-shell form factor $F_1(Q^2,t)$ as a function of the transferred momentum $Q^2$ (left panel) and the off-shell parameter $t$ (right panel) with the results given in~\cite{Choi:2019nvk},~\cite{Leao:2024agy}. The red rhombus shows the results of the model with the constant vertex function ($\Gamma_{\pi}^{(\textrm{CON})}$), the blue triangle shows the results of the model with the symmetric vertex function ($\Gamma_{\pi}^{(\textrm{SYM})}$), and the green triangle shows the results of the present work. The calculated results are close to those of the model with the symmetric vertex function because the vertex functions of the models have a similar structure, unlike the model with the constant vertex function. The results  for the form factor $g(Q^2,t)$ in Figure ~\ref{fig3} for different models are similar to those for the form factor $F_1(Q^2,t)$.

\newpage
\section{Conclusion}

In the present work, the half off-shell EM form factors of the pion $F_1(Q^2,t)$ and $F_2(Q^2,t)$ were calculated
using the Bethe-Salpeter equation with a separable kernel for the vertex function of the pion. 
The parameters of the model were determined in the previous work from the analysis of the pion constants. 
Also, the form factor $g(Q^2,t)$ was obtained that is nonzero on the mass shell and can be extracted
from experimental data.
The half off-shell form factor $F_2$ was calculated in two ways: by direct calculations and by using the Ward-Takahashi identity. The results obtained by these two methods coincide, which implies that the Ward-Takahashi identity is satisfied by the model
presented in this paper.

The investigated model was compared with the results of other authors~\cite{Choi:2019nvk},~\cite{Leao:2024agy}. 
The pion constants ($\sqrt{\langle r_\pi^2\rangle}$, $f_\pi$ and $g(0, m_\pi^2)$) calculated in this work describe the experimental values more accurately. Furthermore, the half off-shell form factors are similar to the results of a model with a symmetric vertex function as opposed to a constant one. It is planned to extend the model to the time-like domain of the square of the transferred momentum in the future as well as consider the radial part of the vertex function's dependence on other functional dependences (dipole and exponent).

\end{document}